\documentclass[aps,prd,twocolumn,superscriptaddress,showkeys,showpacs]{revtex4}
\usepackage{pdfsync}
\usepackage{mathptmx}
\usepackage[utf8]{inputenc}
\usepackage[T1]{fontenc}
\usepackage{slashed}
\usepackage{times}
\usepackage{amssymb,amsfonts,amsmath,amsthm}
\usepackage{dsfont,bbm}
\usepackage{dcolumn}
\usepackage{epsf}
\usepackage{graphicx}
\usepackage[caption=false]{subfig}
\usepackage{dsfont}
\usepackage{slashed}
\usepackage[active]{srcltx}
\usepackage[usenames]{color}

\begin{document}

\title{On $\xi$-process for DVCS-amplitude}

\author{I.~V.~Anikin}
\email{anikin@theor.jinr.ru}
\affiliation{Bogoliubov Laboratory of Theoretical Physics, JINR,
             141980 Dubna, Russia}

\begin{abstract}
In this note, we demonstrate in detail the $\xi$-process implementation applied to the deeply virtual Compton scattering amplitude
to ensure both the QCD and QED gauge invariance. The presented details are also important
for understanding of the contour gauge used in different processes.

\end{abstract}
\date{\today}
\maketitle

\section{Introduction}
\label{Intro}

At present, the role of the two-photon processes, such as
the deeply virtual Compton scattering (DVCS) and the hadron production in two-photon collisions,
in extracting of the information about the generalized parton
distributions (GPDs) and the generalized distribution amplitudes (GDAs) is very well understood in the physical community,
see the review \cite{Belitsky:2005qn} and the references therein.
Both GPDs and GDAs belong to a new kind of parton distributions which give a possibility
to study many subtleties of the composite hadron structure.

We stress that the most of interesting physical effects are related to
the cases where the transversities of different nature present in the considered processes.
For example, the sizable transverse transfer momentum, $\Delta^\perp\not=0$, in the corresponding two-photon
processes leads to the significant single-spin asymmetries where the interference of
leading twist-$2$ and higher twist-$3$ contributions is dominant.
However, the necessity of twist-$3$ contributions reveal the nontrivial problem with the
gauge invariance of the corresponding amplitudes compared to the leading twist-$2$ amplitudes.
Generally speaking, it was clear that in the case of substantial transfer momentum
all sources of the kinematical and dynamical transversities have to be included in the consideration.
However, from the technical point of view this problem remains unsolved till the beginning of
$00$'s when the theoretical solution of the gauge invariance problem appeared in
the DVCS amplitude has been found for the first time in \cite{Anikin:2000em}.
Then, a series of works has been issued, see
\cite{Penttinen:2000dg, Belitsky:2000vx, Vanderhaeghen:2000bt, Radyushkin:2000jy, Belitsky:2000gz, Kivel:2000rb, Brodsky:2000xy},
using the different approaches and extending the results of \cite{Anikin:2000em}.

In \cite{Anikin:2000em}, for the case of deeply virtual Compton scattering off (pseudo)scalar particles
the Ellis-Furmanski-Petronzio-Efremov-Teryaev factorization procedure has been adopted and
the complete expression for the DVCS amplitude up to twist-$3$ contributions has been calculated.
However, the so-called $\xi$-process described in \cite{Bogolyubov:1980nc} which ensures both the QCD and QED
gauge invariance of the DVCS-like amplitude has not been discussed in detail.
Meanwhile, the $\xi$-process plays the unique role not only for the gauge invariance but also for the
deep understanding of the sources of transversity.

In the present paper, we take remedial action and illuminate the important details of the $\xi$-process applied for the
DVCS-like amplitudes which are also useful for the application of the contour gauge.

\section{$\xi$-process}
\label{xi-subpr}

We begin this section with the definition of $\xi$-process applied for an arbitrary amplitude in the Abelian
$U(1)$ gauge theory, QED.
The extension to the non-Abelian $SU(3)$ gauge theory, QCD, is rather trivial and it does demand additional explanations.

Let us consider the amplitude with one photon external line, ${\cal A}_\mu(q, p_i | A)$, where $q$ corresponds to
the photon momentum while $p_i$ denote the momenta that remain after singling out the photon momentum.
The $U(1)$ gauge transformation (gradient transformation) is given by
\begin{eqnarray}
\label{Ph-g}
A^\Lambda_\mu(x) = A_\mu(x) + \partial_\mu \Lambda(x).
\end{eqnarray}
Hence, after performing the gauge transformation of Eqn. (\ref{Ph-g}), the amplitude ${\cal A}_\mu(q, p_i | A)$
is $U(1)$ gauge invariant if the term $\partial_\mu \Lambda(x)$ of Eqn.~(\ref{Ph-g}) does not transform the given amplitude, {\it i.e.}
\begin{eqnarray}
\label{GI-1}
{\cal A}_\mu(q, p_i | A)={\cal A}_\mu(q, p_i | A^\Lambda)\quad \text{if}\quad {\cal A}_\mu(q, p_i | \partial\Lambda)=0.
\end{eqnarray}
Using the $S$-matrix formalism,
\begin{eqnarray}
\label{S-m}
&&S(g)\stackrel{\text{def}}{=}\mathbb{T}\text{exp}\Big\{i\int dx g(x){\cal L}_I(x)\Big\}=
\\
&&
1 +
\sum_n \frac{i^n}{n!} \int dx_1 ... dx_n \, S^{(n)}(x_1,...,x_n)\,g(x_1)...g(x_n),
\nonumber
\end{eqnarray}
the statement of Eqn.~(\ref{GI-1}) takes the following form \cite{Bogolyubov:1980nc}
\begin{eqnarray}
\label{GI-2}
\frac{\partial}{\partial \xi^\mu} \frac{\partial S^{(n)}(x_1,...,x_n)}{\partial A_\mu(\xi)}=0.
\end{eqnarray}

In our case, the set of diagrams with one external photon line can be obtained from the set of diagrams
without the external photon line by means of the insertion of the photon vertex, depending on $\xi$-position, into any external or internal lines.
This insertion is called the $\xi$-process. In other words, the $\xi$-process describes diagrammatically the following correspondence
\begin{eqnarray}
\label{GI-Corr}
\frac{\partial S^{(n)}(x_1,...,x_n)}{\partial A_\mu(\xi)}\Big|_{\xi=x_i} \Longleftrightarrow S^{(n-1)}(x_1,...,x_{n-1}).
\end{eqnarray}

Hence, in order to fulfil the gauge invariance condition, see Eqn.~(\ref{GI-2}),
we first insert the $\xi$-vertex into the external or internal lines and, then, we calculate the divergence over $\xi$-position
to get zero finally.  Notice that for the amplitude written in $p$-representation the 
differentiation over $\xi$-position has been replaced by the contraction with the inserted photon momentum $q$.
If we implement the $\xi$-process only for the internal lines we get the relations reflecting the Ward identity.

\begin{figure}
\centerline{\includegraphics[width=0.4\textwidth]{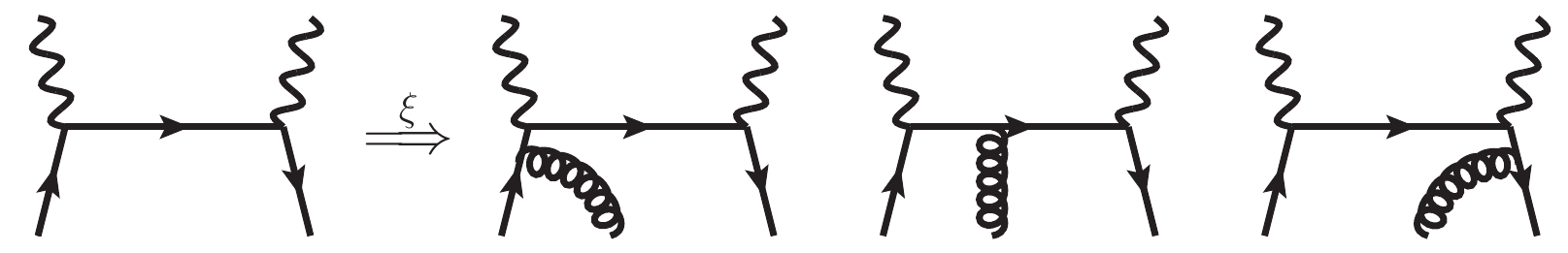}}
\caption{The QCD $\xi$-process for quark DVCS, $\gamma(q)+q(k_1)\to \gamma(q^\prime)+q(k_2)$.
Before $\xi$-process, we deal with the standard diagram involving the quark-photon subprocess.
After $\xi$-process, the first and third diagrams refer to the non-standard diagrams, while the second diagram 
defines the standard diagram with the quark-photon-gluon subprocess.}
\label{Fig-1}
\end{figure}

We are now in a position to consider the $\xi$-process applied for the DVCS parton subprocess, see Fig.~\ref{Fig-1}.
Having applied the QCD $\xi$-process to the deeply virtual Compton scattering off quarks, see Fig.~\ref{Fig-1}
we obtain the following (here, $\hat k = k\cdot\gamma$ and the causality prescriptions are irrelevant for the moment)
\begin{eqnarray}
\label{q-dia-1}
&&{\cal A}^{\mu\nu}_\alpha(\text{dia.-}1)=
\\
&&
\bar u(k_2) \gamma^{\nu} \frac{\hat k_1 -\hat\ell +\hat q}{(k_1-\ell+q)^2} \gamma^\mu
\frac{\hat k_1 -\hat\ell }{(k_1-\ell)^2} \gamma^\alpha u(k_1)
\nonumber
\end{eqnarray}
and
\begin{eqnarray}
\label{q-dia-2}
&&{\cal A}^{\mu\nu}_\alpha(\text{dia.-}2)=
\\
&&
\bar u(k_2) \gamma^{\nu} \frac{\hat k_1 -\hat\ell +\hat q}{(k_1-\ell+q)^2} \gamma^\alpha
\frac{\hat k_1 +\hat q }{(k_1+q)^2} \gamma^\mu u(k_1)
\nonumber
\end{eqnarray}
and
\begin{eqnarray}
\label{q-dia-3}
&&{\cal A}^{\mu\nu}_\alpha(\text{dia.-}3)=
\\
&&
\bar u(k_2) \gamma^{\alpha} \frac{\hat k_2 +\hat\ell}{(k_2+\ell)^2} \gamma^\nu
\frac{\hat k_1 +\hat q}{(k_1+q)^2} \gamma^\mu u(k_1).
\nonumber
\end{eqnarray}
Then, contracting the sum of diagrams with the gluon momentum $\ell$, we derive that
\begin{eqnarray}
\label{ell-dia-sum}
&&\ell_\alpha {\cal A}^{\mu\nu}_\alpha(\text{dia.-}1+\text{dia.-}2+\text{dia.-}3)=
\\
&&
- \bar u(k_2) \gamma^{\nu} \frac{\hat k_1 -\hat\ell +\hat q}{(k_1-\ell+q)^2} \gamma^\mu u(k_1)+
\nonumber\\
&&
\bar u(k_2) \gamma^{\nu} \frac{\hat k_1 -\hat\ell +\hat q}{(k_1-\ell+q)^2}
\Big[ 1- (\hat k_2 + \hat q^\prime)
\frac{\hat k_1 +\hat q }{(k_1+q)^2}\Big]
\gamma^\mu u(k_1) +
\nonumber\\
&&
\bar u(k_2) \gamma^\nu
\frac{\hat k_1 +\hat q}{(k_1+q)^2} \gamma^\mu u(k_1) \equiv 0.
\nonumber
\end{eqnarray}
Thus, Eqn.~(\ref{ell-dia-sum}) demonstrates the QCD gauge invariance of the quark DVCS amplitude.

\begin{figure}
\centerline{\includegraphics[width=0.4\textwidth]{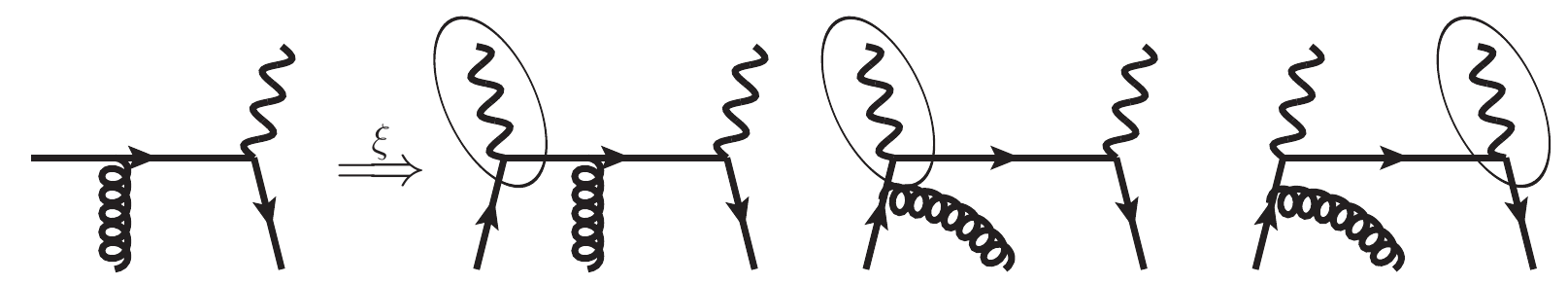}}
\caption{The QED $\xi$-process for quark-gluon-quark-photon process, $q(k_1)+g(-\ell)\to \gamma(q^\prime)+q(k_2)$.
The oval blob denotes the photon which has been inserted.}
\label{Fig-2}
\end{figure}

\begin{figure}
\centerline{\includegraphics[width=0.4\textwidth]{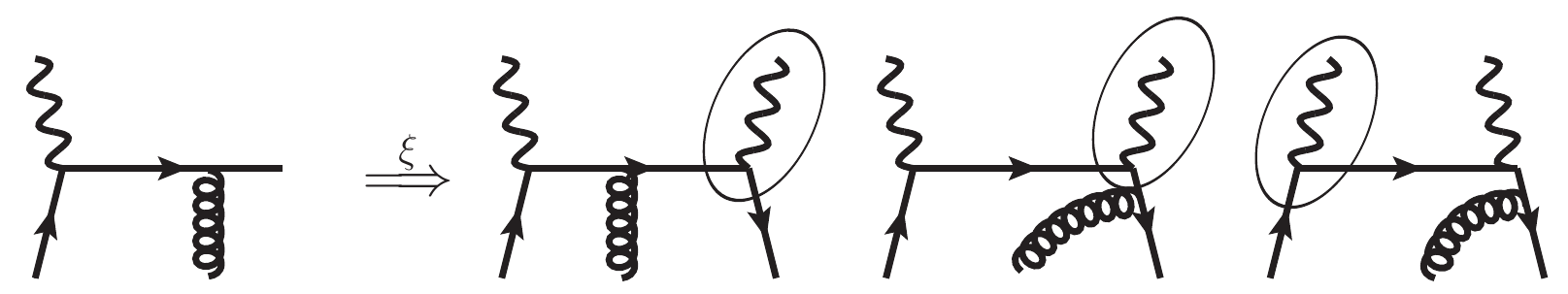}}
\caption{The QED $\xi$-process for quark-photon-quark-gluon process, $\gamma(q)+q(k_1)\to q(k_2)+ g(-\ell)$.
The oval blob denotes the photon which has been inserted.}
\label{Fig-3}
\end{figure}

Also, it is instructive to mention on QED gauge invariance of the quark-gluon-quark photon and
quark-photon-quark-gluon amplitudes for which $\xi$-processes have presented in Figs.~\ref{Fig-2} and \ref{Fig-3}.


\section{Factorization procedure}
\label{Fact-th}

In this section, we remind briefly the main steps of factorization procedure for the DVCS process where we deal with
the substantial hadron momentum transfer, see \cite{Anikin:2000em} for details. The DVCS process is defined as
\begin{eqnarray}
\label{DVCS-pr-1}
\gamma^*(q) + \text{hadron}(p_1) \to \gamma(q^\prime) + \text{hadron}(p_2)
\end{eqnarray}
with
\begin{eqnarray}
\label{DVCS-kin}
\Delta=p_2- p_1, \quad P=(p_2+p_1)/2, \quad
\bar Q=(q+q^\prime)/2.
\end{eqnarray}
Here, the virtuality of the initial off-shell photon defines the so-called
large scale, i.e. $q^2=-Q^2\to\infty$, while the final photon is
on-shell and $q^{\prime\, 2}=0$.
Due to this kinematics, the considered process is a hard exclusive
reaction and its amplitude can be studied within the factorization procedure.
In other words, the hard processes define the asymptotical
regime for the corresponding amplitude which can be related to the the light-cone formalism.
We introduce a  light-cone basis which is constructed by the
``plus'' and ``minus'' vectors as
\begin{eqnarray}
\label{lcb}
&&p=\big(\Lambda,\, 0, \, 0,\, \Lambda\big)=(p^+, 0^-, {\bf 0}_\perp),
\nonumber\\
&&
n=\big(1/(2\Lambda),\, 0, \, 0,\, -1/(2\Lambda)\big)=(0^+, n^-, {\bf 0}_\perp),
\nonumber\\
&&
p\cdot n =1,
\end{eqnarray}
where $\Lambda$ is an arbitrary and dimensionful constant which is
expressed through the Lorentz invariants. The exact form of $\Lambda$
as a function of invariants depends on the frame.
In our case, the vectors $p$ and $n$ are nothing but the average momentum $P$ and the normalized real photon momentum $q^\prime/(P\cdot q^\prime)$,
respectively. In this frame, the relevant vectors can be decomposed in terms of
the Sudakov (light-cone) basis as
\begin{eqnarray}
\label{kin-2}
&&2P=2p+\bar M^2 n \approx \, 2p ,
\nonumber\\
&&\Delta=p_2-p_1 = -2\xi P + 2\xi\bar M^2 n  +\Delta^\perp \approx -2\xi P +\Delta^\perp,
\nonumber\\
&& P\cdot \Delta=0, \quad \Delta^2=t\approx 0,
\end{eqnarray}
where the approximation is valid up to the twist three accuracy, discarding the contributions associated with the twist four and
higher.

Symbolically, the factorization theorem allows the amplitude to be factorized in the form of convolution
as
\begin{eqnarray}
\label{Fac-pr}
\text{Amplitude} = \{\text{Hard (pQCD)}\} \otimes
\{\text{Soft (npQCD)} \}\,.
\end{eqnarray}
In the most ideal case, both the hard and soft parts in Eqn. (\ref{Fac-pr}) are independent of each other,
UV- and IR-renormalizable. Moreover, various parton distributions which parametrize the soft part
have to manifest the universality property.

More exactly, {\it the factorization theorem states that
the short (hard) and long (soft) distance
dynamics can be separated out provided $Q^2$ is large, and the DVCS-like amplitude takes the form of}
\begin{eqnarray}
\label{Con-rep}
&&{\cal A}_{\mu\nu...}=
\int d^4 k_1 d^4 k_2 ... \,\text{tr} \big[E_{\mu\nu...}(k_1,k_2,...) \Phi(k_1, k_2,...) \big]
\nonumber\\
&&
\stackrel{Q^2\to\infty}{ \Rightarrow}
\int dx_1 dx_2 ... \,\text{tr} \big[E_{\mu\nu...}(x_1,x_2,...) \, \Phi(x_1, x_2,...) \big]
\nonumber\\
&&
+
{\cal O}(1/Q^2)
\end{eqnarray}
where $E_{\mu\nu...}$ implies a product of corresponding propagators
which finally forms the hard part of amplitude, and the soft part is related to
the hadron matrix elements of quark-gluon nonlocal operators as
\begin{eqnarray}
&&\Phi(k_1, k_2)\stackrel{{\cal F}}{=} \langle {\cal O}^{(\bar\psi, \psi, A)}(z_1, z_2, 0) \rangle,
\nonumber\\
&&\Phi(x_i) = \int dk^+_i \, \delta(x_i-k_i\cdot n) \,\int dk^-_i\, d^2{\bf k}^T_i \, \Phi(k_i),
\end{eqnarray}
where $\stackrel{{\cal F}}{=}$ denotes the corresponding Fourier transforms defined below, see Eqn.~(\ref{Phi-1}).

The $1/Q^2$-corrections in the hard process amplitudes, see Eqn.~(\ref{Con-rep}), can be classified with the help of
the collinear and(or) geometrical twist defined for the corresponding operators.
The geometrical twist has been determined for local quark-gluon operators as
\begin{eqnarray}
\label{Geo-tw-def}
\tau(\text{twist})=d(\text{dimension})-j(\text{spin}),
\end{eqnarray}
while the collinear twist has been defined for non-local quark-gluon operators as
\begin{eqnarray}
\label{Col-tw-def}
t(\text{coll. twist})=d(\text{dimension})-j_a(\text{spin projection}).
\end{eqnarray}
For instance, in the case of DIS , we have the simplest correspondence
\begin{eqnarray}
\label{DIS-tw-Q2}
\text{Loc.}{\cal O}^{\,\text{twist}=\tau}(\bar\psi, \psi, A) \Longrightarrow
\left(1/Q^2\right)^{\tau/2-1}.
\end{eqnarray}
It is also worth to remind the matching between the contributions of collinear and geometrical twists,
we have
\begin{eqnarray}
\label{mathing}
&&\text{leading twist-} t \Leftrightarrow  \text{leading twist-} \tau
\\
&&\text{next-to-leading twist-} t \Leftrightarrow \tau\leq \text{next-to-leading twist-} t.
\nonumber
\end{eqnarray}
In other words, each of hadron correlators can be presented in the form of
\begin{eqnarray}
\label{matchig-2}
(\text{{L-twist}-}t\,\, \text{operator})
\oplus (\text{NL-twist-}t\,\, \text{operator}) \oplus ....
\end{eqnarray}
where
\begin{eqnarray}
(\text{NL-twist-}t\,\, \text{operator}) \ni (\text{{L-twist}-}\tau\,\, \text{operator}).
\end{eqnarray}

Basically, the factorization procedure (or theorem) gives a recipe for
an asymptotical estimation of the amplitude instead of a explicit calculation.
The latter takes place only if we have defined the hadron-parton interaction (effective) Lagrangian.

In contrast to the Drell-Yan-like (or the Sudakov-like) processes, the DVCS process deals with the only
dominant light-cone direction needed for the factorization procedure. We assume the direction $p^+$, see Eqn.~(\ref{lcb}),
to be a dominant one, {\it i.e.} $p^+ \sim [{\cal P}]\to \infty$.
Therefore, the convolution representation, see Eqn.~(\ref{Con-rep}), is based on the power-counting in according to
\begin{eqnarray}
\label{P-C}
k \sim \big( [{\cal P}], \mu^2/[{\cal P}], \mu \big)\equiv \big( k^+, k^-, {\bf k}_\perp \big)
\end{eqnarray}
applied for all relevant vectors. To get the amplitude factorized, we decompose the
function $E_{\mu\nu...}$ around the dominant direction. This is given by the Taylor expansion as
\begin{eqnarray}
\label{Taylor-expan}
&&E_{\mu\nu...}(k_i) = E_{\mu\nu...}(x_i P) +
\nonumber\\
&&
\frac{\partial E_{\mu\nu...}(k_i)}{\partial k^\alpha_i} \Big|_{k_i=x_iP}\, (k_i-x_i P)^\alpha
+ \ldots,
\end{eqnarray}
together with
\begin{eqnarray}
\label{k}
k_i^\mu =x_i P^\mu + (k_i\cdot P)n^\mu + k^\mu_{i\,\perp} \approx
x_i P^\mu + k^\mu_{i\,\perp}, \quad x_i=k_i\cdot n.
\end{eqnarray}
Notice that the Taylor expansion around the dominant direction, see Eqn.~(\ref{Taylor-expan}),
together with the power-counting, see Eqn.~(\ref{P-C}), lead to the certain constraints for the loop integrations
in the soft part.
Indeed, considering the positive domain of integration for the DVCS amplitude, we have
\begin{eqnarray}
\label{Constr-Soft}
&&{\cal A}_{\mu\nu}= \int_{0}^{\Lambda^\infty} dk^+  E_{\mu\nu}(k^+,0^-, {\bf 0}^\perp)
\nonumber\\
&&
\times\int^{\mu}_{0} d^2 {\bf k}^\perp \,
\int^{\mu^2/\Lambda^\infty}_{0} dk^-\,
\Phi(k^+, k^-, {\bf k}^\perp)\Big|_{k^+=x P^+},
\end{eqnarray}
where $\Lambda^\infty = a [{\cal P}]$.
For the sake of simplicity, we here focus on the one-loop integration and the leading order of expansion.
We can see that in the soft part we have the cuts for the integrations over $dk^-$ and $d^2{\bf k}^\perp$
depending on $\mu$.
Alternatively, instead of the cut-off method we can use the dimension regularization method with the full regions of integration.
We can go over to the full integration with
the integration measure $dk^-\, d^{(D-2)}{\bf k}^\perp$
and, then, we use the RG-equation to study the evolution regarding $\mu$ , {\it i.e.}
\begin{eqnarray}
\label{Int-RG}
&&\int^{\mu}_{0} d^2{\bf k}^\perp \,
\int^{\mu^2/\Lambda^\infty}_{0} dk^- \stackrel{\text{RG}}{\Longleftrightarrow}
\int^{\infty}_{0} d^{(D-2)}{\bf k}^\perp\,
\int^{\infty}_{0} dk^- \Big|_{D=4-2\epsilon}
\nonumber\\
&&\stackrel{\text{def}}{\Longrightarrow}
\int d^2{\bf k}^\perp\, dk^- .
\end{eqnarray}

As a result, we obtain the factorized amplitude in form of mathematical convolution which reads
\begin{eqnarray}
\label{Constr-Soft}
&&{\cal A}_{\mu\nu}= \int_{-1}^{1} dx \, E_{\mu\nu}(x P^+,0^-, {\bf 0}^\perp)
\nonumber\\
&&
\times \int dk^+ \delta (x-k^+/P^+) \int d^2{\bf k}^\perp\,
dk^- \, \Phi(k^+, k^-, {\bf k}^\perp)
\nonumber\\
&&
\equiv
\int_{-1}^{1} dx \, E_{\mu\nu}(x) \,\Phi(x).
\end{eqnarray}
It is instructive to present the soft part of amplitude $\Phi(x)$ through the
Fourier transform of the hadron matrix element of operator. It reads
\begin{eqnarray}
\label{Phi-1}
&&\Phi^{[\Gamma]}(x) = \int (d^4 k) \delta(x-k n)
\nonumber\\
&&
\times \int (d^4 z) e^{i(k-\Delta/2) z}
\langle p_2 | \bar\psi(0) \Gamma \psi (z) | p_1\rangle
\nonumber\\
&&
=\int (d\lambda) e^{i(x+\xi) \lambda}
\langle p_2 | \bar\psi(0) \Gamma \psi (\lambda n) | p_1\rangle
\end{eqnarray}
with the certain Fierz projection defined by $\Gamma$-matrix.
Up to now, our consideration has mainly been focused on the leading term in the Taylor expansion of $E_{\mu\nu}$
and the quark operator related to the soft part of amplitude $\Phi(x)$ up to now.

\section{The soft part of DVCS amplitude with the gluon radiations}
\label{Gluon-R-SF}

We are in a position to discuss the gluon radiations from the quark lines which lead to
the quark-gluon operator (together with the Wilson line) in the function $\Phi$. Since the quark-gluon operators
induce the genuine twist three, we need to take into account the kinematical sources of transversity
related to the presence of the nontrivial $\Delta^\perp$ in the parametrizations.

Sence in the present paper we study the $\xi$-process applied for the amplitude, 
the kinematical twist three can be omitted. The comprehensive analysis of all sources of twist three can be found, for example,
in \cite{Anikin:2011aa, Anikin:2009bf}.

The gluon radiation from the internal quark line, see Fig.~\ref{Fig-1}, can be referred to the standard contribution
of genuine twist three which has been discussed in detail, see \cite{Anikin:2011aa, Anikin:2009bf}. Moreover,
in our previous studies of DVCS-like processes, the role of diagrams with
the gluon radiation from the external quark line of the subprocess has been mentioned rather on the intuitive level.
We rectify the mentioned incompleteness in this paper.

\begin{figure}
\centerline{\includegraphics[width=0.25\textwidth]{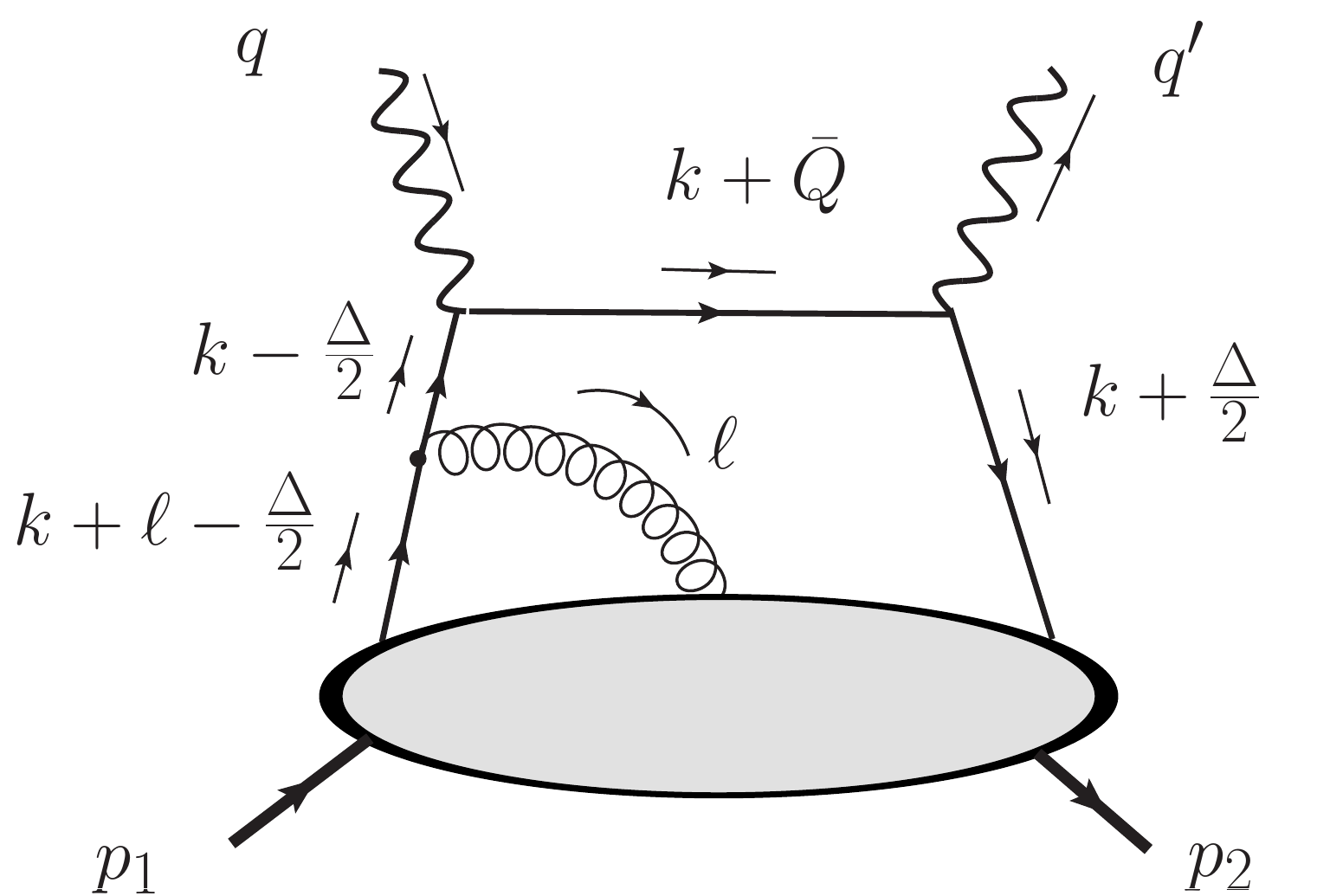}}
\caption{The typical non-standard diagram with the gluon radiation from the external quark line of
subprocess.}
\label{Fig-4}
\end{figure}

Let us dwell on the non-standard diagram
\footnote{The terminology such as non-standard diagrams can be traced from the paper \cite{Anikin:2010wz}.}
generated by the diagram with the gluon
radiation from the external quark line of subprocess, see Fig.~\ref{Fig-4}.
Before the factorization procedure gets applied, we have the following expression for the given amplitude
\begin{eqnarray}
\label{Non-st-1}
{\cal A}_{\mu\nu}(\text{non-stand-1})=
\int (d^4 k) \, \text{tr} \big[ E_{\mu\nu}(k) {\cal F}^{(1)} (k) \big]
\end{eqnarray}
where (by definition, $S(k)=\hat k / (k^2+i\epsilon)$, see Sec.~\ref{xi-subpr}, 
and the prefactor from the massless quark propagator $(-1)/i$ together with the vertex prefactor $i$
have been factorized out and included into the corresponding integration measures)
\begin{eqnarray}
\label{E-1}
E_{\mu\nu}(k) = \gamma_\nu S(k+\bar Q) \gamma_\mu
\end{eqnarray}
and
\begin{eqnarray}
\label{Phi-non-st-1}
&&{\cal F}^{(1)} (k) = S(k-\Delta/2) \gamma_\alpha
\nonumber\\
&&
\times
\int (d^4 z) e^{i(k-\Delta/2)z} \langle p_2|
gA_\alpha(z) \psi(z) \bar\psi(0) | p_1 \rangle.
\end{eqnarray}
According to our factorization procedure, see Sec.~\ref{Fact-th},
we now expand the function $E_{\mu\nu}(k)$ around the dominant direction and we perform the
replacement in the integration measure as
\begin{eqnarray}
\label{subst}
(d^4 k)\Rightarrow (d^4 k) dx \,\delta(x-k n).
\end{eqnarray}
As a result, we have the following factorized amplitude
\begin{eqnarray}
\label{Non-st-1-2}
{\cal A}_{\mu\nu}(\text{non-stand-1})=
\int dx \, \text{tr} \big[ \gamma_\nu S(x P+\bar Q) \gamma_\mu {\cal F}^{(1)} (x) \big]
\end{eqnarray}
where the function ${\cal F}^{(1)} (x)$ reads
\begin{eqnarray}
\label{Phi-non-st-1-2}
&&{\cal F}^{(1)} (x) = \int (d^4 k) \,\delta(x-k n)\,
S(k-\Delta/2) \gamma_\alpha \gamma^-
\nonumber\\
&&
\times
\int (d^4 z) e^{i(k-\Delta/2)z} \langle p_2|
\bar\psi(0) \gamma^+ gA_\alpha(z) \psi(z) | p_1 \rangle
\end{eqnarray}
singling out the vector Fierz projection in the hadron correlator only. Notice that the other Fierz projections, which are irrelevant
for the present study, can be considered in the similar way depending on the cases.

In Eqn.~(\ref{Phi-non-st-1-2}), we concentrate on the twist-$2$ quark combination which is $\bar\psi(0) \gamma^+\psi(z)$.
To get the genuine twist-$3$ quark-gluon operator, we also have to assume $A_\alpha$ to be with the transverse components
which correspond to the collinear twist-$1$ gluon field, $A_\alpha^\perp$.
Hence, the twist-$3$ quark-gluon operator in the hadron correlator leads to the following
\begin{eqnarray}
\label{Phi-non-st-1-3}
&&{\cal F}^{(1)} (x) = \int (d^4 k) \,\delta(x-k n)\,
S(k-\Delta/2) \gamma_\alpha^\perp \gamma^-
\nonumber\\
&&
\times
\int (d^4 z) e^{i(k-\Delta/2)z} \langle p_2|
\bar\psi(0) \gamma^+ gA_\alpha^\perp(z) \psi(z) | p_1 \rangle.
\end{eqnarray}
At the first glance, one can see that nonzero spinor combination corresponds to $S(k-\Delta/2)\sim (k^- -\Delta^-/2)\gamma^+$
which could naively be neglected in accordance with the Sudakov expansion of Eqn.~(\ref{k}).
In fact, the situation is different because we deal with the cancellation of light-cone minus components between
the numerator and denominator of propagator.
Indeed, we have
\begin{eqnarray}
\label{Phi-non-st-1-4}
&&{\cal F}^{(1)} (x) = \int (d^4 k) \,\delta(x-k n) \frac{\gamma^+}{2(k^+ -\frac{\Delta^+}{2})}
\gamma_\alpha^\perp \gamma^-
\nonumber\\
&&\times
\Big[ 1 - \frac{(k^\perp-\frac{\Delta^\perp}{2})^2}{2(k^+ -\frac{\Delta^+}{2})(k^- -\frac{\Delta^-}{2})}
+ O\big( (k^\perp)^2 \big)
\Big]
\nonumber\\
&&
\times
\int (d^4 z) e^{i(k-\Delta/2)z} \langle p_2|
\bar\psi(0) \gamma^+ gA_\alpha^\perp(z) \psi(z) | p_1 \rangle.
\end{eqnarray}
We now parametrize the hadron matrix element of twist-$3$ operator, we have (cf. \cite{Anikin:2000em})
\begin{eqnarray}
\label{Par-tw-3}
&&\int (d^4 z) e^{i(k-\Delta/2)z} \langle p_2|
\bar\psi(0) \gamma^+ gA_\alpha^\perp(z) \psi(z) | p_1 \rangle
\nonumber\\
&&
= P^+\, \Delta^\perp_\alpha\, B\big(k-\Delta/2\big),
\end{eqnarray}
where
\begin{eqnarray}
\int dk^+\delta(k^+ - x P^+) \frac{B\Big(k^+ - \frac{\Delta^+}{2}\Big)}{k^+ -\frac{\Delta^+}{2}}
= \frac{\int dy B(y, x)}{x+\xi}.
\end{eqnarray}

As the next step, we express the quark-gluon para\-me\-tri\-zing function through the quark parametrizing function
using the integral relations which can be stemmed form the QCD equations of motion.
Let us write down the QCD equations of motion
in operator forms (we consider massless quarks only) which reads
\begin{eqnarray}
\label{EOM-1}
\big[\overrightarrow{\hat D(z)} \psi(z)\big] =0 \quad \text{or} \quad
\big[\overrightarrow{\hat D(z)} \psi(z) \bar\psi(0)\big]  = 0
\end{eqnarray}
where all quark and gluon operators are well-defined and act on the suitable Hilbert space.
Going over to the hadron correlator of QCD equations of motion, we write the following
\begin{eqnarray}
\label{EOM-2}
&&\langle p_2|
\overrightarrow{
\hat D(z)} \psi(z) \bar\psi(0) | p_1\rangle = 0,
\nonumber\\
&&
\langle p_2| \psi(z) \bar\psi(0)
\overleftarrow{\hat D(0)}| p_1 \rangle  = 0.
\end{eqnarray}
Notice that the QCD equations of motion, see Eqns.~(\ref{EOM-1}) which are inserted into the hadron correlators, see Eqns.~(\ref{EOM-2}),
with the unphysical (off-shell) hadron states are not needed to be held thanks for the quantum corrections.
However, in our case where the hadron states are described by the physical (on-shell) states, Eqns.~(\ref{EOM-2}) take place.

Using the corresponding parametrizations, these equations lead to the integral relations which can be presented in the form of
\begin{eqnarray}
\label{EOM-3}
&&\int dy \{\text{3-particle GPDs of Fierz proj.}\}(x,y;\xi)=
\nonumber\\
&&\sum_i\{ \text{2-particle GPDs of Fierz proj.}\}_i(x;\xi)
a_i(x,\xi)
\end{eqnarray}
or, more exactly, we have
\begin{eqnarray}
\label{EOM-4}
\int dy B(y, x) = (x+\xi) H_3(x, \xi) + \frac{1}{2} H_1(x, \xi).
\end{eqnarray}
Here, the function $H_1$ and $H_3$ parametrize the hadron correlator of quark (two-partilce) operator as
\begin{eqnarray}
\label{Par-2}
\langle p_2 | \bar\psi(0)\gamma_{\mu} \psi(z) |p_1 \rangle
\stackrel{{\cal F}}{=}
H_1(x, \xi) P_{\mu} + H_3(x, \xi)\Delta^{\perp}_{\mu}.
\end{eqnarray}
The function $H_1$ corresponds to the leading twist-$2$ contribution, while the function $H_3$ is given by the
collinear twist-$3$ combination which involves both the geometrical twist-$2$ (or the kinematical collinear twist-$3$) and
the genuine (or dynamical) twist-$3$. Below, for the $\xi$-process, the kinematical twist-$3$ terms are irrelevant and can be omitted.
Moreover, for the sake of simplicity, we consider only the vector Fierz projection.
The other possible projections can be readily restored by the additive terms \cite{Anikin:2000em}.

\subsection{Contour gauge and Elimination of longitudinal Wilson lines}
\label{C-G}

In conclusion of this Section, we give a short discussion on the use of contour gauge which eliminates
the longitudinal component of gluon field. The necessary details can be found in \cite{Anikin:2016bor}.  
The axial kind of gauges, like $A^+=0$, is in fact a particular case of the most general non-local contour gauge
determined by a Wilson line with a fixed path. For example, the straightforward line
in the Wilson line connecting $\pm\infty$ with $x$ gives us the axial gauge.
It is important that two different contour gauges are able to correspond to the same
local axial gauge.

The contour gauge is a subject of intense studies
(see, for example, \cite{ContourG1, ContourG2}).
The preponderance of contour gauge
is that the quantum gauge theory starts to be free from the Gribov ambiguities.
Moreover, the contour gauge gives the simplest way to fix the residual gauge freedom.
In contrast to the familiar axial gauge, within the contour gauge conception
we first fix an arbitrary point $(x_0, \textbf{g}(x_0))$
in the fiber. Afterwards, we define two directions: one direction is determined in the base
\footnote{The direction in the base $\mathbb{R}^4$ is nothing else than the tangent vector of a curve which
goes through the given point $x_0$.}, the other 
direction -- in the fiber where the direction can be uniquely determined as the
tangent subspace related to the parallel transport. Finally, we can uniquely
define the point in the fiber bundle.

Working with the standard diagram contributions which are generated by the gluon radiations from the 
internal quark line of subprocess, we can derive 
the {\it gauge invariant} quark string operator which takes the form of
\begin{eqnarray}
\label{OP1}
&&\bar\psi(0^+,0^-,{\bf 0}_\perp) [0^+,0^-,{\bf 0}_\perp; \, 0^+,+\infty^-,{\bf 0}_\perp]_{A^+}
\Gamma\times
\\
&&[0^+,+\infty^-,{\bf 0}_\perp; \, 0^+,z^-,{\bf 0}_\perp]_{A^+} \psi(0^+,z^-,{\bf 0}_\perp)
\nonumber\\
&&
=\bar\psi(0^+,0^-,{\bf 0}_\perp) 
\Gamma [0^+,0^-,{\bf 0}_\perp; \, 0^+,z^-,{\bf 0}_\perp]_{A^+} \psi(0^+,z^-,{\bf 0}_\perp),
\nonumber
\end{eqnarray}
where the Wilson line is defined as
\begin{eqnarray}
\label{WL-def}
[z_2 ;\, z_1]_{A}= \mathbb{P}\text{exp}\Big\{ ig \int^{z_2}_{z_1} d\omega_\alpha \,A_{\alpha}(\omega) \Big\}.
\end{eqnarray}
In Eqn.~(\ref{OP1}), $\Gamma$ implies a relevant combination of $\gamma$-matrices.

The non-standard diagram contributions involving the gluon radiations from the
external quark line of subprocess give us the string operator which reads
\begin{eqnarray}
\label{OP2}
&&\bar\psi(0^+,0^-,{\bf 0}_\perp) [-\infty^+,0^-,{\bf 0}_\perp; \, 0^+,0^-,{\bf 0}_\perp]_{A^-}
\Gamma\times
\nonumber\\
&&[0^+,z^-,{\bf 0}_\perp; \, -\infty^+,z^-,{\bf 0}_\perp]_{A^-} \psi(0^+,z^-,{\bf 0}_\perp).
\end{eqnarray}
Following the contour gauge conception,
we now eliminate all the Wilson lines with the longitudinal gluon fields $A^+$ and $A^-$
by the following requirements  
\begin{eqnarray}
\label{C-G-2}
&&[0^+,0^-,{\bf 0}_\perp; \, 0^+,+\infty^-,{\bf 0}_\perp]_{A^+} = \mathds{1},
\nonumber\\
&&[0^+,+\infty^-,{\bf 0}_\perp; \, 0^+,z^-,{\bf 0}_\perp]_{A^+} = \mathds{1}
\end{eqnarray}
and
\begin{eqnarray}
\label{C-G-3}
&&[-\infty^+,0^-,{\bf 0}_\perp; \, 0^+,0^-,{\bf 0}_\perp]_{A^-} = \mathds{1},
\nonumber\\
&&[0^+,z^-,{\bf 0}_\perp; \, -\infty^+,z^-,{\bf 0}_\perp]_{A^-} = \mathds{1}.
\end{eqnarray}

\section{Genuine twist-$3$ contribution and $\xi$-process}
\label{Gen-tw-3}

Having used the results of preceding sections,
we calculate the contribution of the diagrams where the gluon radiation has been related
with the (anti)quark fields which enter the hadron matrix element. We have
\begin{eqnarray}
\label{H3-add-1}
&&{\cal A}^{\mu\nu}(\text{non-stand.-}1)=
-\frac{1}{2} \int_{-1}^{+1} dx \, \frac{H^{(g)}_3(x, \xi)}{x-\xi +i\epsilon}
\nonumber\\
&&
\times \Big\{
-\big[ (xP+\bar Q)_\nu \Delta^\perp_\mu + (\mu\leftrightarrow \nu) \big] +
\nonumber\\
&&
P\cdot \bar Q \big[ \Delta^\perp_\nu n_\mu - \Delta^\perp_\mu n_\nu \big]+
(x-\xi) \big[ \Delta^\perp_\mu P_\nu - \Delta^\perp_\nu P_\mu \big]
\Big\}
\end{eqnarray}
and
\begin{eqnarray}
\label{H3-add-2}
&&{\cal A}^{\mu\nu}(\text{non-stand.-}2)=
-\frac{1}{2} \int_{-1}^{+1} dx \,\frac{H^{(g)}_3(x, \xi)}{x-\xi +i\epsilon}
\nonumber\\
&&
\times \Big\{
-\big[ (xP+\bar Q)_\nu \Delta^\perp_\mu + (\mu\leftrightarrow \nu) \big] -
\nonumber\\
&&
P\cdot \bar Q \big[ \Delta^\perp_\nu n_\mu - \Delta^\perp_\mu n_\nu \big]-
(x-\xi) \big[ \Delta^\perp_\mu P_\nu - \Delta^\perp_\nu P_\mu \big]
\Big\}.
\end{eqnarray}
Then, after summation we obtain 
\begin{eqnarray}
\label{H3-DVCS}
&&{\cal A}^{\mu\nu}(\text{non-stand.-}1)+{\cal A}^{\mu\nu}(\text{non-stand.-}2)=
\nonumber\\
&&\int_{-1}^{+1} dx \,\frac{H^{(g)}_3(x, \xi)}{x-\xi +i\epsilon}
\Big\{
(xP+\bar Q)_\nu \Delta^\perp_\mu + (\mu\leftrightarrow \nu) \Big\}
\nonumber\\
&&
\equiv {\cal A}^{\mu\nu}(\text{standard dia. with the $q-\gamma$ subpr.}).
\end{eqnarray}
Thus, Eqn.~(\ref{H3-DVCS}) shows that the non-standard diagram contributions, see Fig.~\ref{Fig-1},
which involve the parametrizing function $H_3$ are identical to
the standard diagram contribution with the same parametrizing function $H_3$ 
which correspond to the quark-photon subprocess provided the integral relations
reflecting the QCD equations of motion have been used.
In other words, for the QED gauge invariance of the DVCS amplitude it is enough to include the genuine twist-$3$ contribution of
the standard diagram involving the quark-photon subprocess 
together with the standard diagram with the quark-gluon-photon subprocess.

\section{Conclusions}
\label{Conc}

One of the most interesting physical effects is related to
the cases with the sizeable transversities of the hard processes.
In DVCS-like processes, the substantial transfer momentum with $\Delta^\perp\not=0$
can be associated with the several single-spin asymmetries where the interference of
leading twist-$2$ and higher twist-$3$ contributions is dominant.
On the other hand, the necessity of twist-$3$ contributions reveal the nontrivial problem with the
gauge invariance of the corresponding amplitudes compared to the leading twist-$2$ amplitudes.
It was clear that for such kind of processes it is mandatory to study all sources of the kinematical and dynamical transversities.
From the technical point of view, the gauge invariance problem
has been resolved for the DVCS amplitude for the first time \cite{Anikin:2000em}. However, the full analysis
of the $\xi$-process in order to ensure the gauge invariance has not been presented.

In \cite{Anikin:2000em}, the role of the gluon radiation from the external quark line of the subprocess has been hidden.
This sort of diagrams can be referred to the non-standard diagrams in contrast to the
standard diagrams where the gluon has radiated from the internal quark line of the subprocess.
Meanwhile, the non-standard diagrams form the corresponding Wilson lines which are important for the contour gauge
conception, see \cite{Anikin:2010wz}.

In the present paper, we have rectified the mentioned incompleteness.
Having used the integral relations which has been induced by the QCD equations of motion,
we have demonstrated that the non-standard diagram contributions, see Fig.~\ref{Fig-1}, with the function $H_3$ are identical to
the standard diagram contribution with the same function $H_3$ which involves the quark-photon subprocess.
Hence, for the QED gauge invariance of the DVCS amplitude it is enough to include the genuine twist-$3$ contribution of
the standard diagram involving the quark-photon subprocess together with the standard diagram with the quark-photon-gluon subprocess. Moreover, we have shown that the non-standard diagram contributions play the important role to use the contour gauge for the gluon field.

\section*{Acknowledgements}
We are grateful to M.~Deka, D.~Ivanov, N.~Kivel, A.~Manashov, B.~Pire, M.~Polyakov, L.~Szymanowski, O.~Teryaev and S.~Wallon for useful discussions.

\end{document}